\documentclass[repreint,10pt,letterpaper,twocolumn,titlepage,oneside]{revtex4-1}
\usepackage[latin1]{inputenc}
\usepackage{amsmath}
\usepackage{amsfonts}
\usepackage{amssymb}
\usepackage{caption}
\usepackage{color}
\usepackage[english]{babel}
\usepackage{blindtext}
\usepackage{enumerate}
\usepackage{graphicx}
\usepackage{epstopdf}
\usepackage{subcaption}
\begin{document}
\title{The Reach of Threshold-Corrected Dark QCD}
\author{Jayden L. Newstead}
\email{jnewstea@asu.edu}
\author{Russell H. TerBeek}
\email{rterbeek@asu.edu}
\affiliation{Department of Physics, Arizona State University, Tempe, AZ 85287-1504}
\date{May 2014}
\begin{abstract}
We consider a recently-proposed model which posits the existence of composite dark matter, wherein dark ``quarks" transforming as fundamentals under an $SU(3)_d$ gauge group undergo a confining phase and form dark baryons.  The model attempts to explain both the $\mathcal{O}(1)$ relic density ratio, $\Omega_{\mathrm{dark}}/\Omega_{\mathrm{baryon}}\sim 5.4$, as well as the asymmetric production of both dark and baryonic matter via leptogenesis.  Though the solution of $\beta$ functions for $SU(3)_c$ and $SU(3)_d$ constitutes the main drive of the model, no threshold corrections were taken into account as the renormalization scale crosses the mass threshold of the heavy new fields in the model.  We extend this work by explicitly calculating the threshold-corrected renormalization-group flow for the theory using an effective-field matching technique.  We find that the theory has a much wider range of applicability than previously thought, and that a significant fraction of models (defined by the number of fields contained therein) is able to account for the observed relic density.
\end{abstract}
\maketitle
\section{Introduction}
While the basic makeup of the Universe is well known, only 5\% is composed of known particles. The remaining portion is 27\% dark matter (DM) and 68\% dark energy~\cite{Ade:2013zuv}; however, the labeling of these latter components elucidates the extent of our understanding of them. We know the DM exists because of its gravitational phenomena, observed across a wide range of length scales, yet it cannot be observed directly. The success of the standard model (SM) of particle physics behooves us to attempt to fit the DM into this paradigm. By necessity, the particles which make up the DM must (at most) interact weakly with the SM particles, or else experimental efforts to detect them would have already been successful.\\
\indent There are two core ideas for how the DM was produced in the early universe. The dominant paradigm is that of a thermally-produced weakly-interacting massive particle (WIMP). The WIMP miracle supposes that a particle of weak scale mass and interaction strength would freeze out of the thermal bath and annihilate sufficiently to leave behind the correct relic density.  Since ordinary matter has an observed baryon/anti-baryon asymmetry, a more nuanced production mechanism is thus required to account for this fact. Given the distinct origins of the dark and ordinary matter, one has no reason to expect their densities to be of the same order of magnitude. Herein lies the second paradigm: given the similarity of the densities, it is not unreasonable to suppose that the DM may have a common origin with ordinary matter. A broad class of such models, known as asymmetric dark matter (ADM) models, attempt to simultaneously solve the baryon asymmetry problem while tying the abundance of DM to that of ordinary matter (for a recent review of ADM see~\cite{Petraki:2013wwa,Zurek:2013wia}). In doing so the dark sector also becomes asymmetric; in some implementations, this is achieved by hiding the $B-L$ charge in. In recent years a plethora of mechanisms for generating a baryon and dark matter asymmetry have been developed, e.g.,~co-genesis~\cite{Cheung:2011if,MarchRussell:2011fi}, darkogenesis~\cite{Shelton:2010ta}, and hylogenesis~\cite{Davoudiasl:2010am}.  More precisely, each mechanism ties the number-density of DM and ordinary particles together, generally up to an~$\mathcal{O}(1)$~factor (this is not necessarily true in co-genesis models). These models therefore require the DM particle masses to be $\sim1$-$10$~GeV, but, they often offer no explanation for why this scale should also be the same order as the proton mass.\\
\indent The scale of the proton mass is effectively set by the QCD confinement scale; thus, in order to justify a commensurate mass of the ADM particle, Bai and Schwaller posited that there are related strong dynamics in the dark sector~\cite{Bai:2013xga}. In this scenario, the DM particle is the lightest stable hadron of a new `dark' SU(3) group which is coupled to QCD above some scale $M$. This model can explain the baryon asymmetry through a leptogensis mechanism; to obtain the correct ratio of energy densities, one must then have the dark confinement scale $\Lambda_{\mathrm{dQCD}}\sim\Lambda_{\mathrm{QCD}}$. This can be naturally achieved by decoupling the heavy bi-fundamental fields at some scale $M$ (of the terascale or beyond) and requiring this scale to be at the infrared fixed point (IRFP) of both decoupled beta functions. Below $M$ only low-mass quarks transforming fundamentally under either $SU(3)_c$ or $SU(3)_d$ influence the running of the beta function.  We do not attempt here to re-derive the formalism employed in~\cite{Bai:2013xga}; we refer the interested reader to that paper for a detailed examination of the leptogenesis-induced baryogenesis mechanism and further implications for LHC phenomenology.\\
\indent In this paper we aim to extend the analysis of~\cite{Bai:2013xga} by including threshold corrections to the running of the beta function. Utilizing the space of all possible models (characterized by their field content and mass scale), we constrain the available parameter space by requiring the running of~$\alpha_c$~to fit the latest experimental data. We then show that the conclusions of~\cite{Bai:2013xga} are in fact one limiting case of the dark-QCD concept, with a much broader range of applicability.
\section{Two-Loop $\beta$-Functions}
To the second order in the loop expansion, the $\beta$-function for $SU(3)_c$ is given by
\begin{widetext}
\begin{eqnarray}\label{betafunc}
\beta_c & = & \frac{g_c^3}{16\pi^2}\left\lbrack\frac{4}{3}T(R_f)(n_{f_c}+N_d n_{f_j})+\frac{1}{3}T(R_s)(n_{s_c}+N_d n_{s_j})-\frac{11}{3}C_2(G_c)\right\rbrack\nonumber\\
& +&\frac{g_c^5}{(16\pi^2)^2}\Big\lbrack\left(\frac{10} {3}C_2(G_c)+2C_2(R_f)\right)T(R_f)2(n_{f_c}+N_d n_{f_j})\nonumber\\
& +&\left(\frac{2}{3}C_2(G_c)+4C_2(R_s)\right)T(R_s)(n_{s_c}+N_dn_{s_j})-\frac{34}{3}C_2^2(G_c))\Big\rbrack\nonumber\\
& +&\frac{g_c^3g_d^2}{(16\pi^2)}\lbrack 2C_2(R_f)T(R_f)2N_dn_{f_j}+4C_2(R_s)T(R_s)N_dn_{s_j}\rbrack,
\end{eqnarray}
\end{widetext}
with the analogous $\beta$-function for the dark coupling $g_d$ obtained by swapping the indices $c\leftrightarrow d$.  As in \cite{Bai:2013xga}, the group theory factors are included for complete generality, but we ultimately want the matter fields to transform as fundamentals under $SU(N)_c\times SU(N)_d$~for~$N=3$, and the gauge fields to transform as adjoints.  The trace over generators in the fundamental representation is given by $T(R_f)=T(R_s)=1/2$; the quadratic Casimirs of the adjoint representation are $C_2(G_{c,d})=N_{c,d}$, and those of the fundamental representation are $C_2(R_f)=C_2(R_s)=(N_{c,d}^2-1)/(2N_{c,d})$.
%
%
%
%
%
%
%
\section{Methods} \label{methods}
A given model is completely characterized by its field content:
\begin{equation}\label{fields}
\{n_{f_c},\,n_{f_{c,h}},n_{f_d},\,n_{f_j},\,n_{s_c},\,n_{s_d},\,n_{s_j}\},
\end{equation}
where the indices respectively enumerate the light colored quarks (always 6, as in the Standard Model), heavy colored quarks (transforming under\,$SU(3)_c$\,just as conventional quarks do, but with significantly higher masses), dark quarks, bi-fundamental quarks, colored scalars, dark scalars, and bi-fundamental scalars.  Bi-fundamental fields transform fundamentally under the product group~$SU(3)_c\times SU(3)_d$, as the name suggests.  For clarity, we use the phrasing~\emph{e.g.}~``the field index~$n_{f_d}$" rather than ``the number of dark-fundamental quarks~$n_{f_d}$."  In order to populate all possible models for analysis, we begin by solving the coupled~$\beta$~functions of Eqn.\,\eqref{betafunc} while setting all elements of Eqn.\,\eqref{fields} to zero, save $n_{f_{c,h}}$.  We then determine the maximal field index $n_{f_{c,h}}^{\mathrm{max}}$ that would keep the one-loop term of Eqn.\,\eqref{betafunc} negative; if this upper bound were exceeded, the resulting theory could not reasonably be called QCD.  Next, we set $n_{f_{c,h}}=0$, and vary $n_{f_j}$ in likewise fashion.  After iterating through the possible index values of all fields transforming non trivially under $SU(3)_c$, we then switch to $\beta_d$ and repeat the process with fields fundamental under $SU(3)_d$.  This procedure creates approximately 32 million models for further study.  Although they meet the most minimal definition of an acceptable QCD-like theory, further cuts are necessary in order to produce models capable of a GeV-scale DM candidate.  Next, we require that the models satisfy:
\begin{enumerate}
	\item $\beta_{c,d}\leq 0$ for all energies below the IRFP;
	\item The value of the color fine structure constant ($\alpha_c$) may not 
	exceed the experimentally-observed value measured at the highest 
	energy scale currently available; at the time of writing, the CMS 
	collaboration has measured 
	$\alpha_c(896\,\mathrm{GeV})=0.0889 \pm 0.0034$~\cite{Chatrchyan:2013txa};
	\item The value of the dark fine structure constant ($\alpha_d$) may not 
	exceed the Cornwall-Jackiw-Tomboulis bound \cite{Cornwall:1974vz} 
	of $\pi/4$ when evaluated at the IRFP.
\end{enumerate}
Upon applying this set of constraints, the number of candidate models reduces to 87,286.  Given a set of models with this minimal consistency, it is now possible to begin applying matching conditions, as discussed in \cite{Chetyrkin:1997un}, from the low-energy effective field theory (EFT) (in this case, just the SM) to the full theory.\\
\indent In MS-like renormalization schemes, the Appelquist-Carazzone decoupling theorem \cite{Appelquist:1974tg} does not generally apply to quantities that do not represent physical observables; examples include coupling constants and~$\beta$~functions in the absence of ultraviolet completions.  This means that heavy fields circulating in loops do not necessarily decouple from such quantities at energies below their mass scale.  The solution is to use the EFT formalism, as in \cite{Chetyrkin:1997un}, and elaborated in~\cite{Martens:2010nm,Bauer:2008bj}.  In their original studies, Chetyrkin \emph{et~al.}~\cite{Chetyrkin:1997un}~considered, among other things, the decoupling relations that result from integrating out a single quark flavor from an $n_f$-flavor theory, and requiring consistency with the resulting $(n_f-1)$-flavor EFT that accurately describes the dynamics at lower energies.  This consistency condition is
\begin{equation}\label{matching_condition}
\alpha_c^{\mathrm{EFT}}(\mu_0)=\zeta_c^2\,\alpha_c(\mu_0),
\end{equation}
where\,$\mu_0$ is the IRFP scale to be solved for, and the decoupling function for a single heavy quark flavor at one-loop order is
\begin{equation} \label{zeta_1}
\zeta_c^2=1-\frac{\alpha_c(\mu)}{6\pi}\mathrm{ln}\left(\frac{\mu^2}{M^2}\right).
\end{equation}
The work of \cite{Chetyrkin:1997un} extends to $\mathcal{O}(\alpha_c^3)$; and at this order in perturbation theory, one must take into account the scheme-dependence of the decoupling function, for instance in MS scheme vs.\;on-shell scheme.  However, since the~$\beta$~functions we use extend to two-loop order, it is only consistent to consider threshold corrections at first order in perturbation theory, and at this order all schemes agree.  Furthermore, since we are considering all heavy fields to possess the same mass scale $M$, it follows that Eqn.\,(\ref{zeta_1})\;ought to be replaced by
\begin{equation} \label{zeta_N}
\zeta_c^2=1-\frac{\alpha_c(\mu)}{6\pi}\Big\lbrack n_{f_{c,h}}+N_dn_{f_j}+\frac{1}{4}(n_{s_c}+N_dn_{s,j})\Big\rbrack\,\mathrm{ln}\left(\frac{\mu^2}{M^2}\right).
\end{equation}
When Eqn.\,(\ref{matching_condition})\;is satisfied, one obtains information about the ratio $\mu_0/M$, rather than\,$M$\,itself.  This fact stands in contrast to \citep{Bai:2013xga}, where, in the absence of threshold effects, the low-energy $\alpha_c$ could be evolved directly up to the fixed point determined by solving $\beta_c=\beta_d=0$, and the resulting energy scale could then be extracted.  This approximation uniquley determines\,$M$\,for a given model.  However, when computing the full evolution of the strong couplings to two-loop order and incorporating the necessary threshold/decoupling effects, we find that\,$M$\,is not specified from first principles, and therefore one must consider a continuum of\,$M$\,values.  When sampling the range $m_{\mathrm{top}}\leq M\leq 100\,\mathrm{TeV}$, we find that the solutions obtained in \citep{Bai:2013xga} are limiting cases of composite DM models with a much broader range of applicability.\\
\indent In this work, once we specify a model that meets our previously-stated constraints, we randomly select an\,$M$\,uniformly distributed along a logarithmic scale running from $m_{\mathrm{top}}$ (taken from~\cite{Beringer:1900zz}) to 100 TeV.  This lower limit is set by the non-detection of any novel, fundamental QCD states at energies less than or equal to the top quark mass.  Although the upper limit is arbitrary, the distribution of models tends to peak around~$M\sim$~a few TeV (as seen in Fig. 2), and we find that pushing the upper limit much further does not significantly alter the proportion of models with GeV-scale DM candidates.\\
\indent Once\,$M$\,is fixed, $\mu_0$ may be uniquely determined.  One may then independently evolve\;$g_c,\,g_d$\;down from the decoupling scale to lower values.  Whereas a detailed calculation of the dynamics of chiral symmetry breaking and the resulting hadronization would require a comprehensive lattice study, we follow \citep{Bai:2013xga} in adopting the Cornwall-Jackiw-Tomboulis condition for chiral symmetry breaking\,\citep{Cornwall:1974vz}: when $\alpha_d(\mu) C_2(R_f)=\pi/3$, or equivalently $\alpha_d(\mu)=\pi/4$, the $\mu$ at which this occurs is identified with the chiral symmetry-breaking scale $\Lambda$.  Working with the low-energy QCD of the Standard Model, one finds the relationship $m_p\approx 1.5\,\Lambda$.  We therefore apply the same relation to learn the approximate dark proton mass, $m_d$, from the chiral symmetry-breaking scale~$\Lambda_{\mathrm{dQCD}}$~of dark QCD.\\
\indent Given that the decoupling scale $\mu_0$ may now be somewhat lower than the scale $M$, one might inadvertently modify the running of $\alpha_c$ at observable scales. We thus employ a $\chi^2$ test to restrict ourselves to models in which measured values of $\alpha_c$ are not greatly affected. Using data from CMS (the 31 highest energy data points from~\cite{Chatrchyan:2013txa}) we calculate the figure of merit
\begin{equation}
t = \Sigma_i\frac{(x_i-\alpha_{c,i})^2}{\sigma_i^2},
\end{equation}
where the $x_i$ represent the values of $\alpha_c$ we obtain at the energy scale of the $i$th measurement. The quantity $t$ follows a $\chi^2$ distribution; using this, we calculate a p-value and reject unsatisfactory models at the 95\% confidence level.
\section{Results}
\indent After performing 20 iterations of the above test per model, and indexing through all models that pass our three criteria, we eventually narrow down the list of candidate solutions from 87,286 to 16,859.  It must be noted that this refinement follows from the artificial upper bound of 100 TeV which we set on the universal mass\,$M$\,of all heavy fields.  If the upper limit were to be increased further, a model which did not make this most recent cut would have a broader range of~$M$~values from which to sample, and one could then find a~$\mu_0$~satisfying~$0<\mu_0<M$. \\
\indent When we solve for the fixed-point couplings,\;$\alpha_{c,d}^*$, we find that there is no correlation whatsoever between the magnitude of the coupling and the precise distribution of field content across the available indices.  In fact, for a given number of degrees of freedom\;$N$\;spread across different possible field indices,\;$\alpha_{c,d}^*$\;takes on all values in the allowed interval.  Even though\;$N$\;and\;$\alpha^*$\;are unrelated, there is an argument for the ability of 100 TeV-scale particles to influence GeV-scale physics.  There exist some models for which\;$\alpha_c^*$\;is very small, which necessarily means that the coupling of the EFT must be evolved to very large\;$\mu_0$\;in order to meet it.  So, to first order in Eqn.\,(\ref{zeta_N}), one sets $\alpha^{\mathrm{EFT}}_c(\mu_0)=\alpha_c^*$ for very large\;$\mu_0$.  To next order, one must then satisfy
\begin{equation}
\alpha^{\mathrm{EFT}}_c(\mu_0)=-\frac{(\alpha_c^*)^2}{6\pi}\,N\,\mathrm{ln}(\mu_0^2/M^2),
\end{equation}
where~$N$~sums over all degrees of freedom that couple to QCD, and still subject to the requirement $\mu_0<M$.  Taking $\alpha_c^{\mathrm{EFT}}(\mu_0)=\alpha_c^*$ from the previous approximation results in
\begin{equation}\label{mu_hierarchy}
\mu_0=M\,\mathrm{exp}\,\left(-\frac{3\pi}{N\alpha_c^*}\right).
\end{equation}
Provided that $\alpha_c^*\lesssim 3\pi N^{-1}$, Eqn.\,(\ref{mu_hierarchy}) predicts a natural hierarchy of scales between $\mu_0$ and $M$.  Once SM- and dark-QCD decouple at the scale $\mu_0$, the evolution of their couplings depends solely on the number of light degrees of freedom transforming fundamentally under one group only.  This means that, for $\mathrm{100\,TeV}\simeq M\gg\mu_0$, a light baryon-like state can still be achieved even if just a few ($n_{f_d}\simeq 3$) matter fields contribute to the beta function, thereby causing\;$\alpha_d$\;to run more slowly and trigger chiral symmetry breaking at a lower energy scale.  To recapitulate,
\begin{enumerate}
\item Low\;$\alpha^*_c$ implies large\;$\mu_0$;
\item For\;$\alpha_c^*\lesssim 3\pi\,N^{-1}$, one finds $\mu_0\ll M$, allowing for the relevance of very massive fields;
\item Sufficient $n_{f_d}$ (from 1 to 3) can evolve $\alpha_d^*$ slowly from the IRFP value at the high scale to the Cornwall-Jackiw-Tomboulis bound of $\pi/4$ at the comparatively low GeV scale.
\end{enumerate}
\indent Below we show the plots for the allowed field content of dark QCD.  The blue bars of each histogram indicate the field indices for which the models passed the three criteria outlined in the Section\,\ref{methods}, whereas the violet bars indicate the field indices corresponding to models for which
\begin{equation}\label{omega}
\frac{1}{3}\left(\frac{\Omega_d}{\Omega_b}\right)_{\mathrm{obs}}\;<\;\frac{\Omega_d}{\Omega_b}\;<\;3\left(\frac{\Omega_d}{\Omega_b}\right)_{\mathrm{obs}},
\end{equation}
where $\left(\Omega_d/\Omega_b\right)_{\mathrm{obs}}$ indicates the observed value quoted from the PLANCK collaboration.  This factor of three encapsulates in an approximate sense the spread of models that are ``near" to the actual relic density ratio observed in the universe today, and how general the reach of dark QCD is in producing candidates that could faithfully replicate the matter density observed in the universe.  Though the criterion of Eqn.\,(\ref{omega}) is not the most precise experimental constraint that could be devised, a more rigorous standard would neglect the fact that the onset of chiral symmetry breaking is a truly non-perturbative phenomenon --- whereas we have used a perturbative approximation to derive\;$\Lambda_{\mathrm{dQCD}}$ --- and hence introduces a larger source of error.\\
\indent When we apply Eqn.\,(\ref{omega}) to the remaining models, we find that 2,578 make the cut, or approximately $15.3\%$ - an~$\mathcal{O}(1)$~fraction of the total.  This suggests an interesting and novel result: for a very diverse range of~$M$,~$\mu_0$, and field indices, the dark QCD theory is capable of producing a DM critical density commensurate with the observed value.  Since this result is independent of such idiosyncracies as a precise combination of colored, dark, and bi-fundamental fields, or a narrow range of the~$M$~parameter, we are left to conclude that the mechanism itself --- a confining~$SU(3)_d$~gauge group that shares a fixed point with~$SU(3)_c$~ --- has some real explanatory reach.  The generality of our results bolsters the idea that composite, strongly-interacting DM can provide a natural explanation for the $\mathcal{O}(1)$ ratio of dark and baryonic matter densities observed in the universe.
\begin{figure*}[h]
\centering
\captionsetup{justification=centering,margin=2cm}
\makebox[\textwidth][s]{
		\includegraphics[scale=0.50]{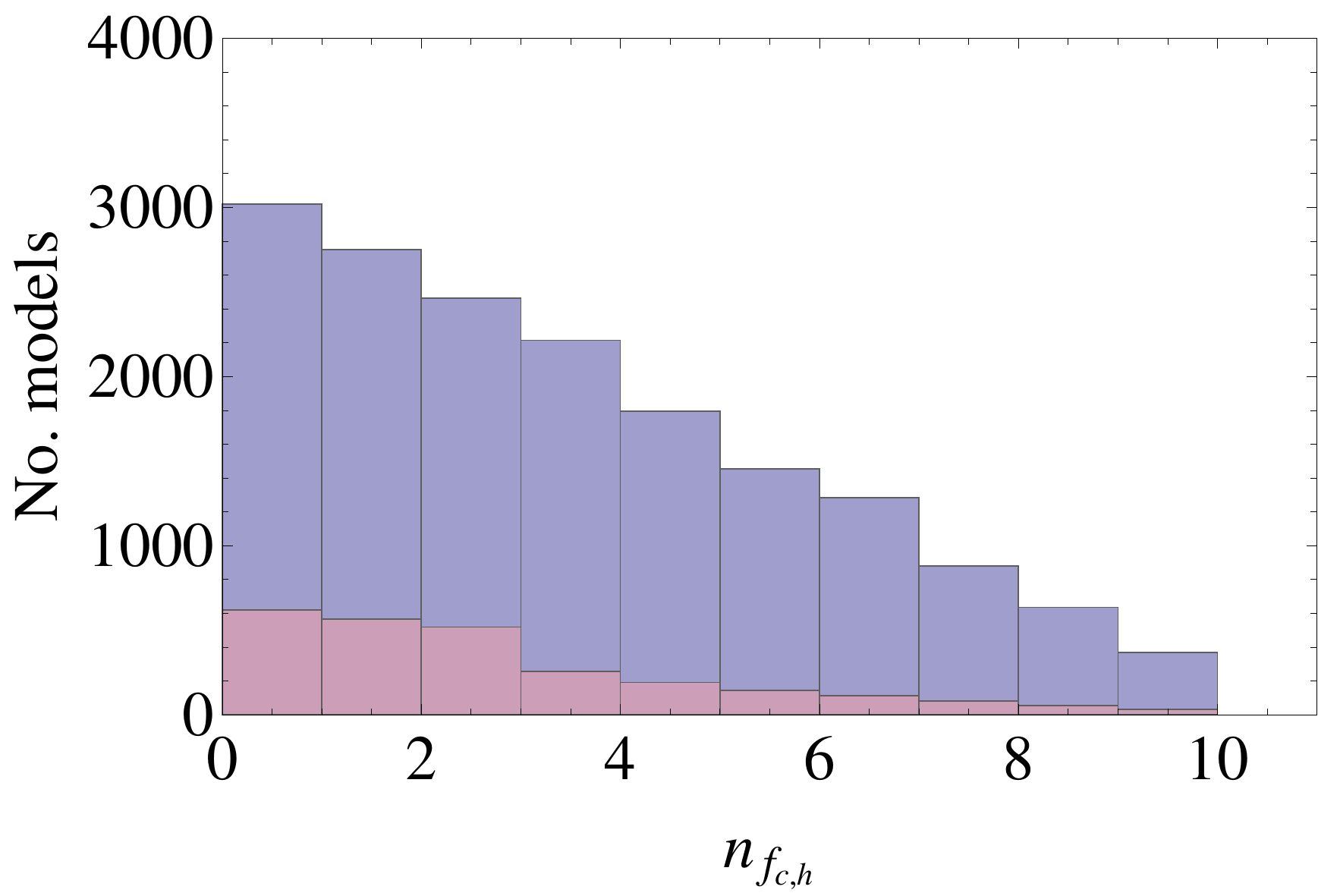}
		\includegraphics[scale=0.50]{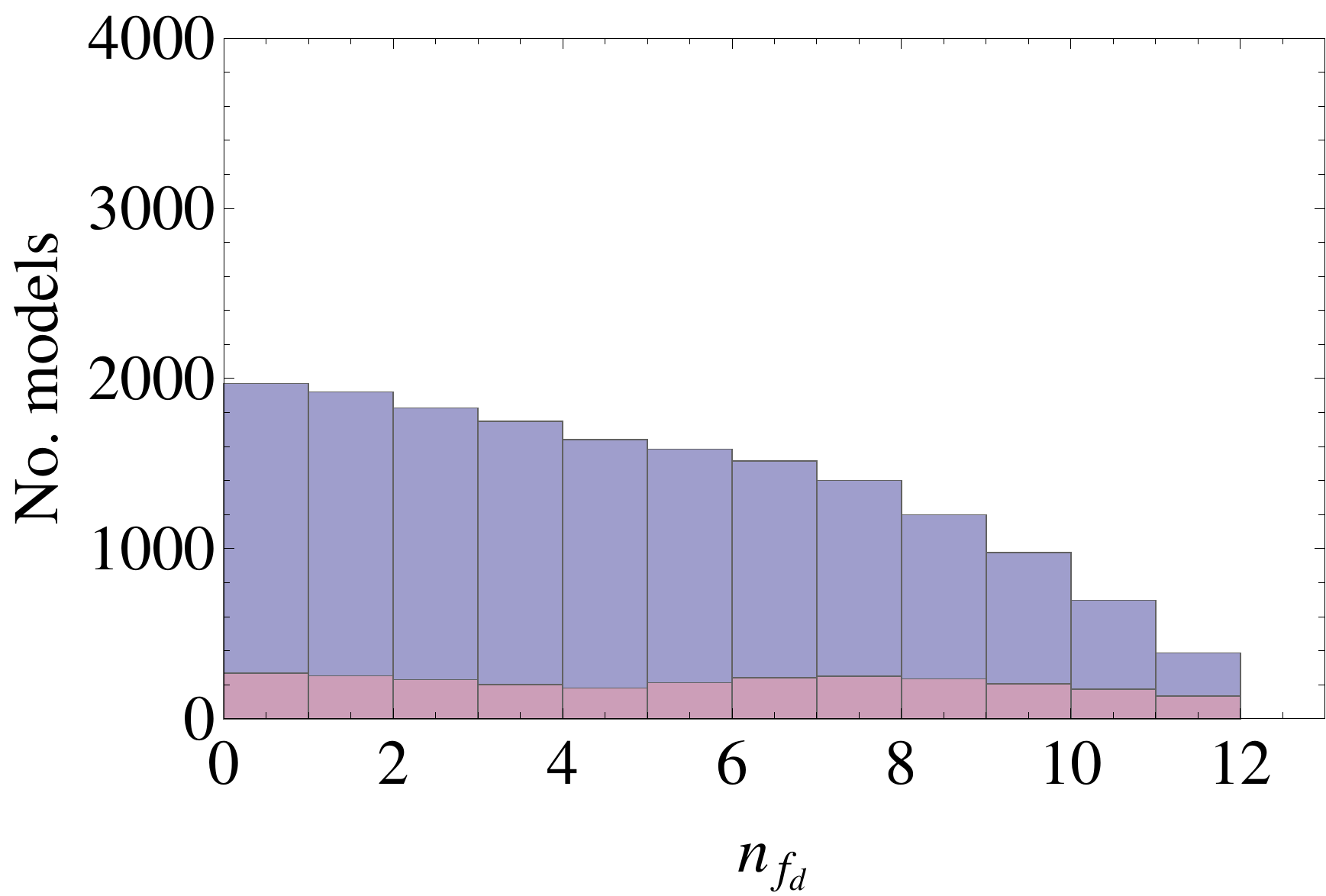}
	}

\makebox[\textwidth][s]{
		\includegraphics[scale=0.50]{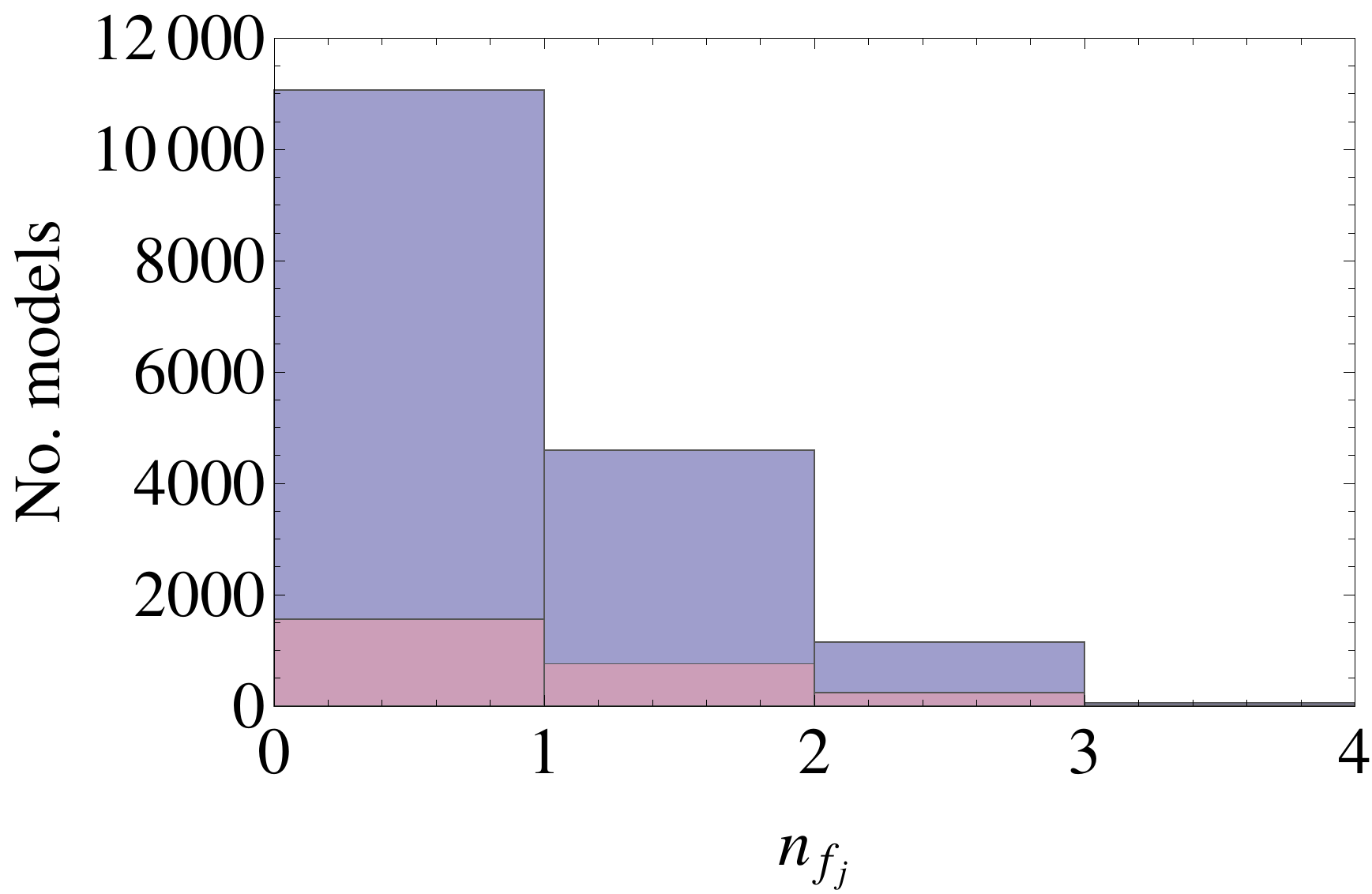}
		\includegraphics[scale=0.50]{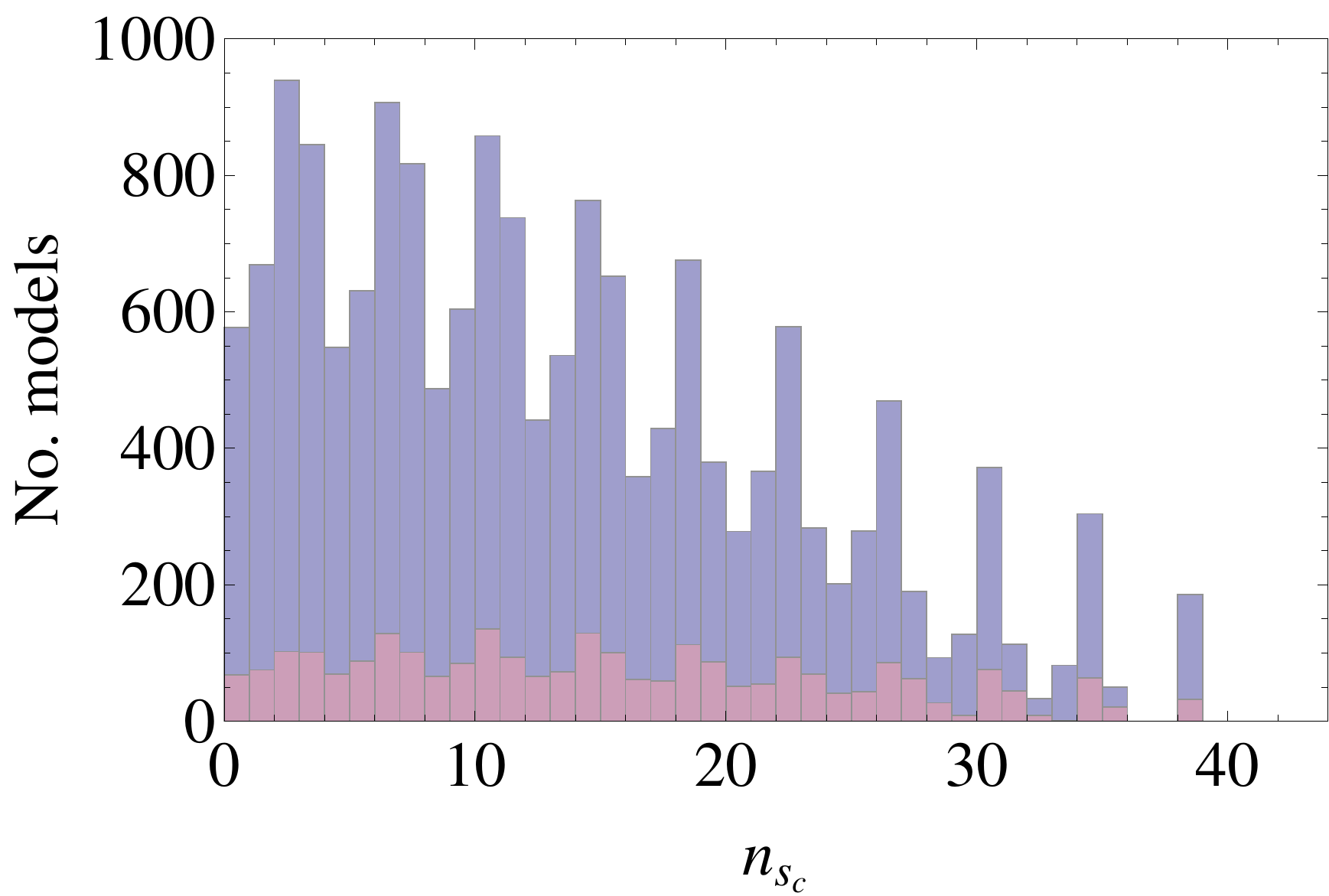}
	}

\makebox[\textwidth][s]{
		\includegraphics[scale=0.50]{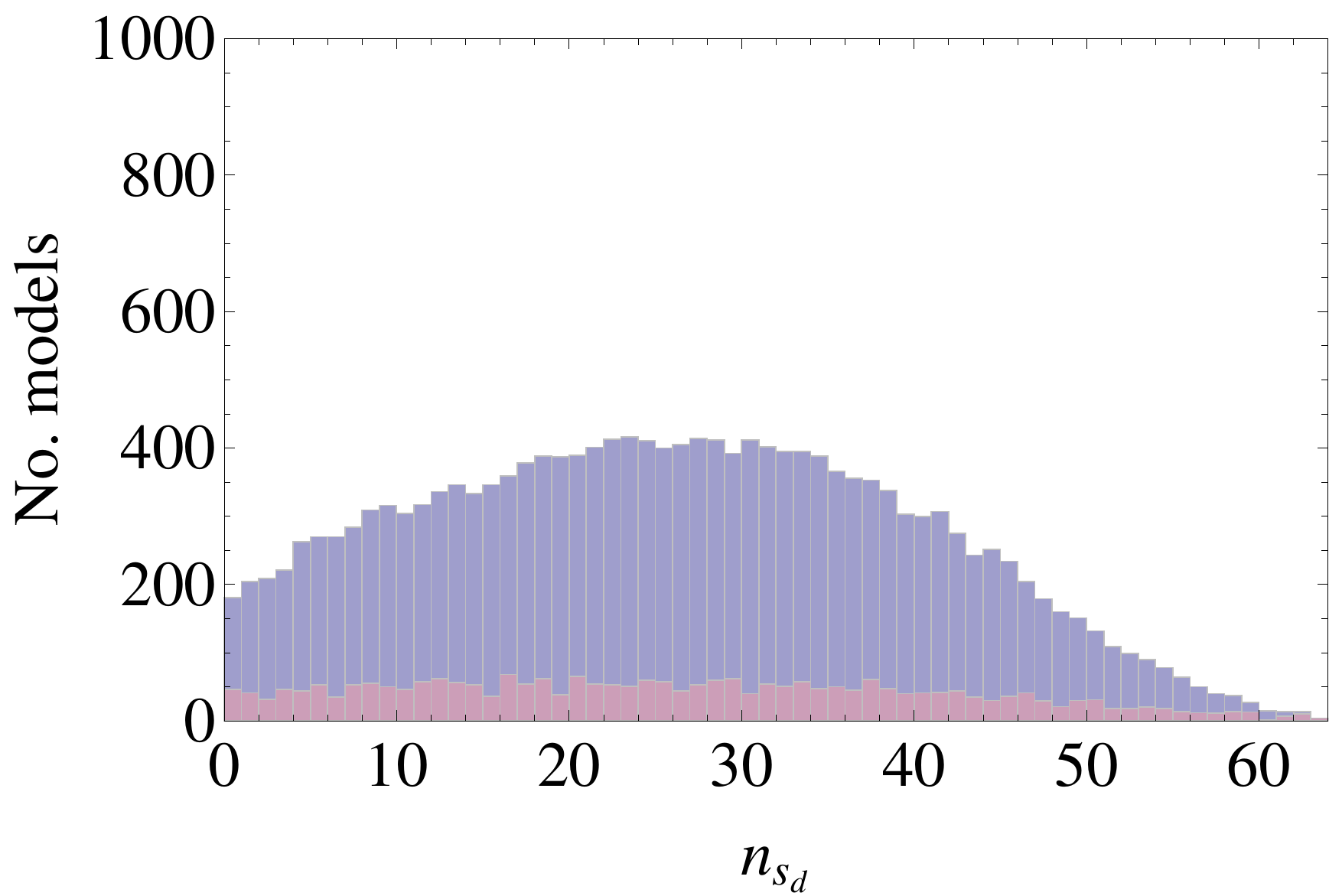}
		\includegraphics[scale=0.50]{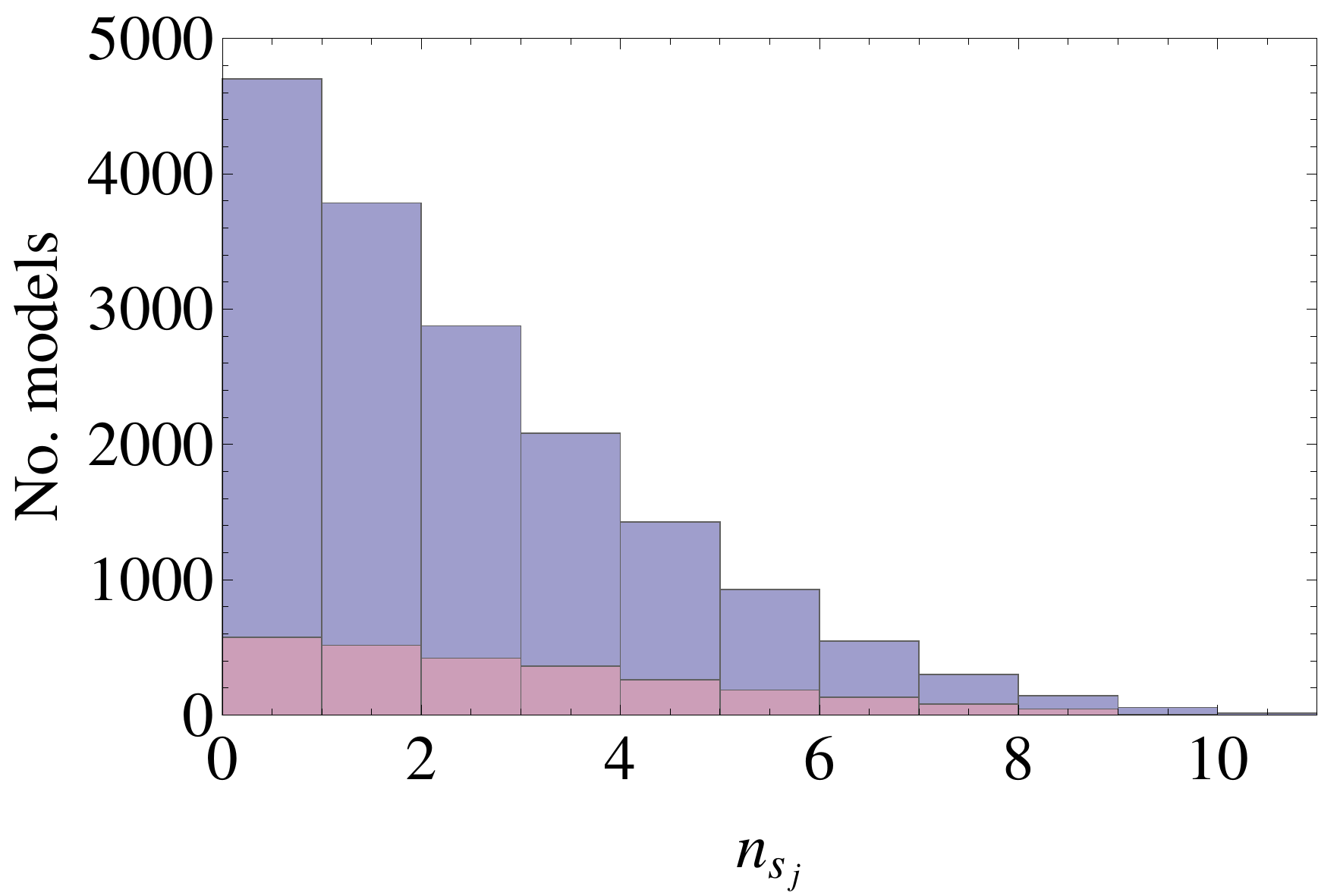}
	}
		\label{index_plots}
		\caption{Number of models which pass our three criteria (blue), 
		as well as the subset that comes within a factor of 3 of $(\Omega_d/
		\Omega_b)_{\mathrm{obs}}$ (violet), organized by field index.}
\end{figure*}
\begin{figure*} \label{fig:M_and_mu_plots}
\captionsetup{justification=centering,margin=2cm}
\centering
	\makebox[\textwidth][s]{
		\includegraphics[width=0.5\textwidth]{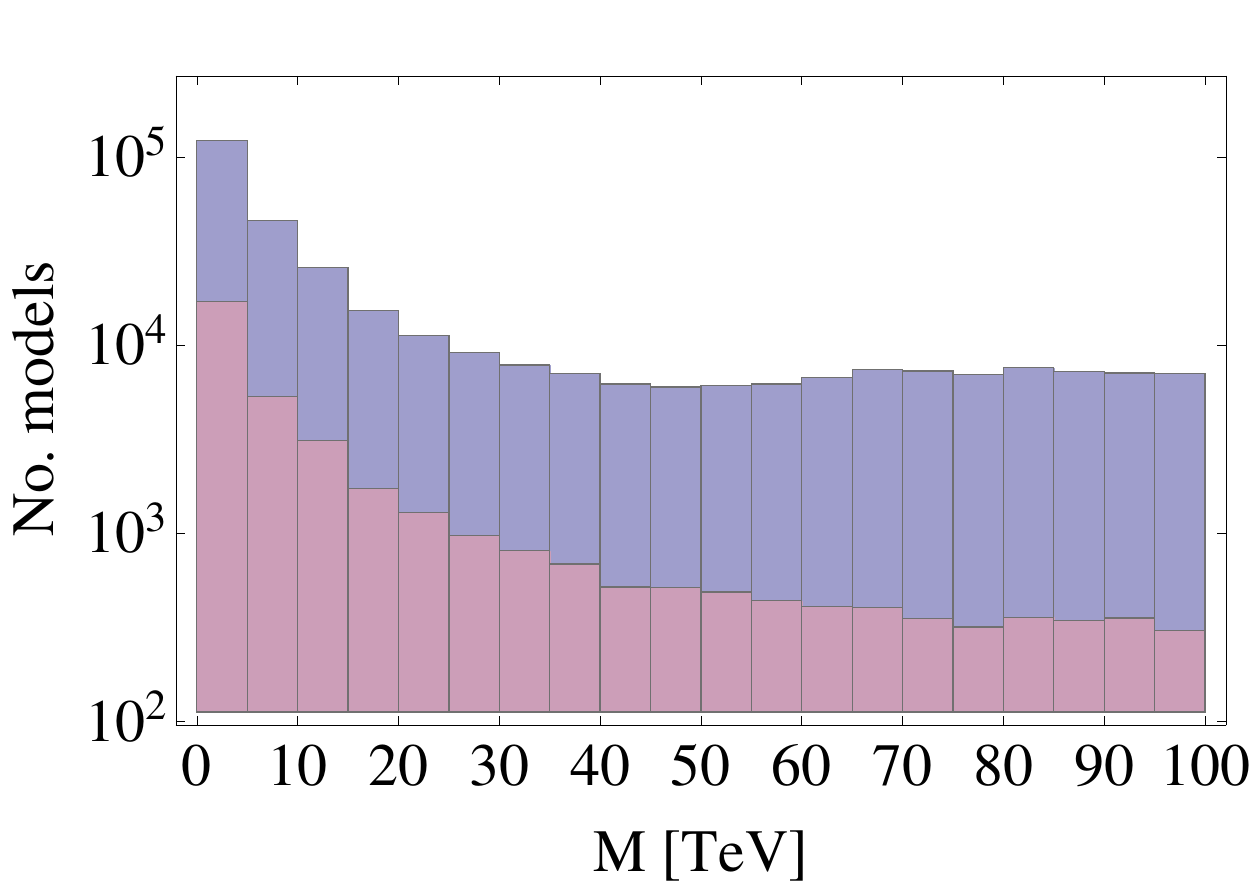}
		\includegraphics[width=0.5\textwidth]{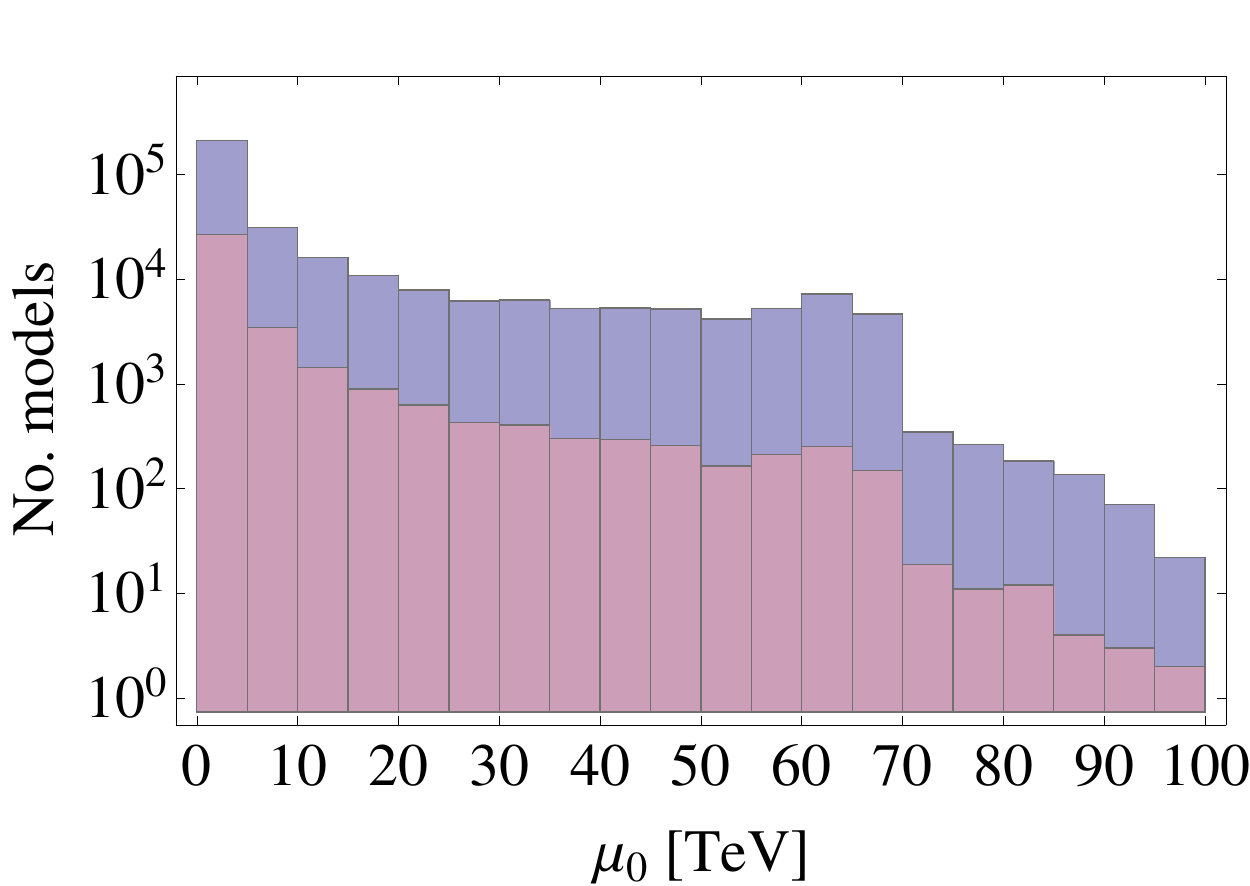}
	}
	\caption{Number of models which pass our three criteria (blue), 
		as well as the subset that comes within a factor of 3 of $(\Omega_d/
		\Omega_b)_{\mathrm{obs}}$ (violet), organized by the mass of the 
		heavy fields\;$M$\;and the decoupling scale\;$\mu_0$.}
\end{figure*}
%
%
%
%
%
%
\\
\section{Conclusions}
The composition and interactions of dark matter remain a mystery to physics, as does the origin of the matter-antimatter asymmetry.  ADM models provide a compelling and elegant way to simultaneously create a matter-antimatter asymmetry and a dark matter component with roughly equal number densities.  We have examined in detail one such model, the $SU(3)_c\times SU(3)_d$ theory of \citep{Bai:2013xga}, in which the relic dark matter candidate is a composite state of ``quarks" transforming fundamentally under $SU(3)_d$, and trivially under $SU(3)_c$.  Our work has extended the scope of this ADM model in order to include decoupling effects due to heavy fields that influence $\beta_c,\;\beta_d$ at very high energies.  By taking decoupling effects into account at one-loop order, we can relax the restriction $M=\mu_0$ implicit in \cite{Bai:2013xga}, thereby opening up a much broader range of possible mass values for the candidate fields to sample.  We find the remarkable result that thousands of models within the dark-QCD framework are able to produce relic density ratios close to that observed in our own universe.  We are therefore led to conclude that ADM models of this type --- $SU(3)_c\times SU(3)_d$ gauge group with an infrared fixed point --- can account for the matter density quite naturally, with a broad diversity of different field types involved.
%
%
%
\begin{acknowledgments}
This research was funded in part by the Department of Energy under Grant No.\,DE-SC0008016 (J.\,L.\,N.) and the National Science Foundation under Grant PHY-1068286 (R.\,H.\,T.).  The authors would also like to thank R. F. Lebed for helpful discussions during the course of this project.
\end{acknowledgments}
\bibliographystyle{plain}
\bibliography{NT2014}
\end{document}